\documentclass[% reprint,
 superscriptaddress,
nofootinbib,
 amsmath,amssymb,
 twocolumn,
 aps,
 pra,
 10pt,
longbibliography
]{revtex4-2}

\usepackage{float}
\usepackage{graphicx,xcolor}
\usepackage{enumitem}
\usepackage{braket}
\usepackage[colorlinks=true]{hyperref}
\usepackage{amsfonts,amssymb}
\usepackage{amsmath, mathtools}
\usepackage{caption,subcaption}

\usepackage[normalem]{ulem}

\usepackage{float}
\usepackage{graphicx,xcolor}
\usepackage{tikz}
\usetikzlibrary{arrows.meta}

\usetikzlibrary{positioning}
\usepackage[edges]{forest}
\usepackage{enumitem}
\usepackage{braket}
\usepackage{comment}

\usepackage{amsthm}
\usepackage{tikz-cd}

\definecolor{amethyst}{rgb}{0.6, 0.4, 0.8}

\begin{document}
\title{Quantum Darwinism and the quality of Petz recovery}

\author{Juha Torvinen}
\email{juha.o.torvinen@helsinki.fi}
\address{Department of Physics, P.O.Box 64, FIN-00014 University of Helsinki, Finland}

\author{Esko Keski-Vakkuri}
\email{esko.keski-vakkuri@helsinki.fi}
\address{Department of Physics, P.O.Box 64, FIN-00014 University of Helsinki, Finland}
\address{InstituteQ - the Finnish Quantum Institute, Finland}
\address{Helsinki Institute of Physics, P.O.Box 64, FIN-00014 University of Helsinki, Finland}

\author{Nicola Pranzini}
\email{nicola.pranzini@helsinki.fi}
\address{Department of Physics, P.O.Box 64, FIN-00014 University of Helsinki, Finland}
\address{InstituteQ - the Finnish Quantum Institute, Finland}
\address{QTF Centre of Excellence, Department of Physics, University of Helsinki, Helsinki, Finland}

\begin{abstract}
   According to Quantum Darwinism, system-environment interactions both \textit{einselect} particular system properties and \textit{encode} them redundantly in many independent subsets of the environment, called fragments. This redundancy implies that an observer can recover the einselected information by accessing just one such fragment. However, the protocol by which such reconstruction should occur is often left unspecified. Considering a system $\Gamma$ interacting with a multipartite environment $\Xi$, we investigate whether, and under what conditions, the einselected state of $\Gamma$ can be recovered from environmental fragments using the Petz recovery map. We show that the fidelity between the system's initial state and the state reconstructed via Petz recovery develops a plateau as a function of the fragment size. Our results are supported by both analytical arguments and numerical simulations of large but tractable models.
\end{abstract}

\maketitle

\section{Introduction}
A quantum system interacting with an environment composed of many subsystems undergoes decoherence, a kind of dynamics that suppresses quantum features such as superpositions and coherences~\cite{BreuerP02}. Due to inhibition of quantum features, this process has long been proposed as the mechanism responsible for making quantum systems appear classical~\cite{Zeh70, Zurek03, Schlosshauer07}. In practice, a state undergoing decoherence will lose most of its quantum properties, while the features that survive the interaction are said to be einselected~\cite{Zurek81, Zurek82, Zurek98}. Beyond mapping quantum states to einselected classical ones, a model for classical emergence must also endow quantum systems with objective properties. Classicality requires more than just classical-looking dynamics and traits, as the system’s observables must also become objective in the classical regime, for different observers should be able to measure any observable associated with the einselected pointer basis, in any order, and obtain consistent results~\cite{Zurek09}. This stands in stark contrast to standard quantum measurements, which inevitably disturb the system and prevent this type of objectivity from arising~\cite{Schlosshauer05}. In recent years, several frameworks have been developed to explain how objectivity might emerge from quantum mechanics: quantum Darwinism (QD)~\cite{Zurek00, Zurek03, Zurek09}, Spectrum Broadcast Structures~\cite{KorbiczEtAl14, HorodeckiEtAl15}, and strong quantum Darwinism~\cite{LeOC19}. Their relationships were clarified in \cite{Korbicz21}. A central idea shared by these approaches is that observers access the system through the environment, which acts as an intermediary by redundantly encoding information about the system. Objectivity is then analysed in terms of the structure and accessibility of this environmental encoding~\cite{Zurek22}. More specifically, quantum Darwinism proposes that classicality emerges from the information about a system that is (i) robust under interaction with a large, uncontrollable environment and (ii) redundantly encoded across many environmental fragments through that interaction. Once the einselected information about the system is encoded in the environment in this way, multiple observers can independently access different fragments of the environment and therefrom infer the same properties of the system. Due to this redundant encoding across many parts of the environment, it is then argued that different observers accessing separate fragments of the environment can independently infer the same classically selected state of the system. While the dynamical aspect of the above picture is well studied and understood, the decoding step is much less discussed in the literature (to the authors' best knowledge). Yet, understanding the precise operation by which an observer can retrieve information about the system by interacting only with fragments of the environment is crucial for explaining how classical, objective properties operationally emerge from underlying quantum ones.

In practice, QD is usually associated with the emergence of a redundancy (or classical) plateau. Consider a system $\Gamma$ interacting with a multipartite environment $\Xi$. We call a fragment any collection of environmental subsystems, and introduce an ordered family of fragments such that each element is larger than all the previous ones, and the largest fragment is $\Xi$ itself. After some physically motivated unitary evolution is applied, the system will generally decohere as it gets correlated with the environment~\cite{Schlosshauer07}. We may then compute the (mutual) information shared between the system and each fragment in the family, seeing it as a function of the label parametrising the latter. The interaction is said to exhibit a \textit{redundancy plateau} if the plot of this function displays an extended and nearly constant region (bounded by an initial and a final ramp), at a value typically close to the entropy of the reduced state of $\Gamma$ after the interaction, under conditions of strong decoherence and efficient environmental encoding~\cite{Zurek09}. The appearance of a redundancy plateau then indicates that the interaction distributes information about the system broadly and uniformly throughout the environment, redundantly encoding the classical, einselected aspects of its state therein. Note that while QD is sufficiently generic to account for the emergence of objectivity in realistic settings, it is most easily benchmarked when the system-environment interaction displays permutation invariance over the environmental parts~\cite{ChiribellaDA06, RiedelZ10}; the redundancy plateau becomes independent of the choice of fragment family in this case. More generally, some notion of typicality of system-fragment interaction is often sufficient for QD to emerge~\cite{OllivierEtAl04, Branda0EtAl15}, and models where permutation invariance is relaxed in this way are as old as QD itself~\cite{BlumeKohoutZ05, BlumeKohoutEtAl08, ZwolakEtAl09, Perez10}.

In practice, the presence of a redundancy plateau indicates that the information surviving the interaction is distributed throughout the environment in such a way that different observers, each accessing different fragments, should be able to reconstruct the same state for $\Gamma$. These reconstructed states need not coincide with the initial one, but they should contain the same einselected information and match with each other. In other words, an observer with access to only a portion of the environment should be able to objectively recover the classicalised features of the encoded state, independently of which fragment is accessed and, to some extent, of its size. Although this intuition is widespread in the literature, the precise meaning of ``reconstruct'' and the mechanism by which the information decoding should occur is seldom made explicit. This article addresses this gap by formalising the decoding process using the tools of recoverability in quantum information theory~\cite{Wilde15, Wilde17}. We employ the Petz recovery map, a mathematical tool describing a physically implementable channel reversing the effects of quantum processes under suitable conditions~\cite{FawziR15}. Petz recovery maps have been recently employed in many fields, from open system dynamics~\cite{LautenbacherEtAl22, LautenbacherEtAl24, KwonM22}, quantum thermodynamics~\cite{KwonK19}, holography~\cite{CotlerEtAl19, ChenEtAl20, AlmheiriEtAl15, BahiruV23, FuruyaLM22}, and error correction in quantum computing~\cite{BarnumK02, NgM10, BiswasEtAl24, MandayamN10} and, more generally, in any von Neumann algebra~\cite{FuruyaLO2022}. By simulating the decoherent, measurement-like evolution of a given initial state, and applying the Petz map, we can evaluate the effectiveness of recovery as a tool of environmental fragment decoding. In this case, we obtain that the recovery is efficient in correspondence with a redundancy plateau. Yet, we also found that objectivisation and classicalisation are not required for a good recovery. Our theoretical findings are supported by numerical simulations.

\section{Tools}
This article examines whether, and how, the information required to reconstruct a system's state is redundantly encoded in fragments of the environment the system is interacting with. Considering a dynamics that exhibits typical properties of quantum Darwinism, we examine (i) if it encodes the principal system's state in environmental fragments, and (ii) how this encoding is distributed across the environment. The encoding we consider is characterised as the possibility of (at least partially) recovering the principal system's state by acting on the environmental fragments with an appropriate quantum channel. 

To set the stage for our analysis, we combine insights from open‑systems dynamics and decoherence with recoverability techniques from quantum information theory. Hereafter, any quantum system $\Psi$ has a state space denoted by $\mathcal{D}(\Psi)$; this is the set of positive, trace‑class operators with unit trace acting on a complex separable Hilbert space $\mathcal{H}_\Psi$.

\subsection{Einselection}
\label{s.QD}
We study the encoding of quantum information in the environment by a premeasurement-like interaction (\textit{a.k.a.} von Neumann measurement). The reason for this choice is that, even if describing a specific kind of interaction, it models system-environment interactions in a way that is generic enough to find broad applicability in many contexts. Dynamically, this interaction is realised by the Ozawa coupling as follows~\cite{Ozawa23, BuschEtAl96}. Consider a principal system $\Gamma$ in the state $\rho_\Gamma$ and let it interact with an environment $\Xi$ in the initial state $\ket{\Xi_R}$ via the Hamiltonian 
\begin{equation}
    H_{int}=O_\Gamma\otimes O_\Xi~.
    \label{e.H_generic}
\end{equation}
Let us assume $O_\Gamma$ is non-degenerate, and diagonalise it as
\begin{equation}
    O_\Gamma=\sum_r \omega_r\ket{r}\bra{r}~;
\end{equation}
then, we can express the time evolution operator as the conditional operation:
\begin{equation}
    U(t)=e^{-itH_{int}}=\sum_r \ket{r}\bra{r}\otimes U^r(t)~,
\end{equation}
with 
\begin{equation}
    U^r(t)=e^{-it\omega_r O_\Xi}~.
\end{equation}
Using the pointer basis $\{\ket{r}\}$ to express the initial state of the system,
\begin{equation}
    \rho_\Gamma=\sum_{r,r'}\rho_{r,r'}\ket{r}\bra{r'}~,
    \label{e.Gamma_init}
\end{equation}
we get that the initial separable state $\rho_\Gamma\otimes\ket{\Xi_R}\bra{\Xi_R}$ gets mapped into the joint state
\begin{equation}
    \rho(t)=\sum_{r,r'}\rho_{r,r'}\ket{r}\bra{r'}\otimes\ket{\Xi^r(t)}\bra{\Xi^{r'}(t)}
\end{equation}
with $\ket{\Xi^r(t)}=U^r(t)\ket{\Xi_R}$.

While easy to show, it is not directly evident from the above that the interaction ``selects'' the pointer states as a privileged subset of states of the principal system. To show that this is indeed the case, consider the state of the principal system after the interaction
\begin{equation}
    \rho_\Gamma(t)={\rm Tr}_\Xi[\rho(t)]=\sum_{r,r'}\rho_{r,r'}\braket{\Xi^r(t)|\Xi^{r'}(t)}\ket{r}\bra{r'}
\end{equation}
and compare it with the initial state \eqref{e.Gamma_init}. If we want the state to be invariant under the interaction, we need $\braket{\Xi^r(t)|\Xi^{r'}(t)}=1$ for all $r,r'$ such that $\rho_{r,r'}\neq 0$. This is, of course, trivial when $r=r'$, but impossible to achieve at all times for all $r\neq r'$~\cite{HeinosaariZ11}. Hence, convex combinations of pointer states are the only states invariant under the effects of the environment. Under very general conditions, it is also possible to show that $\braket{\Xi^r(t)|\Xi^{r'}(t)}\to 0$ for large $t$ and $r\neq r'$ holds for realistic environments (\textit{i.e.}, large and disordered ones, see \textit{e.g.} \cite{Zurek82, PranziniV24}). Adding this ingredient, the above also implies that any non-pointer state will evolve into a mixture of pointer states after long enough time. In this sense, pointer states are (ein-)selected by the system-environment interaction. 

Note that we neglected any free evolution of the subsystems $\Gamma$ and $\Xi$ in the above discussion. This choice was made not to obscure the details of the interaction and subsequent principal system-to-environment encoding.

\subsection{Quantum Darwinism}
We now turn our attention to whether and how information about the system flows into the environment during einselection. Given that the Hamiltonian \eqref{e.H_generic} models a measurement-like interaction, it is perhaps not surprising that, after the coupling, the environment acquires information about the system. Yet, the Hilbert-space localization properties of how this information is encoded across subsystems are nontrivial. To study this, one typically examines the information shared between the principal system $\Gamma$ and fragments of the environment $F_k$, where each fragment comprises a collection of environmental subsystems~\cite{Zurek09}. The information shared between these is quantified by the quantum mutual information $I(\Gamma : F_k)=S(\Gamma)+S(F_k)-S(\Gamma\cup F_k)$, where $S(X)$ denotes the von Neumann entropy of the system $X$. Specifically, one studies how the mutual information varies with the size and localisation of the fragment within the environment. Limiting our study to cases where the system–environment state exhibits an effective permutation invariance among the environmental subsystems, only the fragment’s size is expected to play a role. When permutation invariance holds, all fragments of the same size are equivalent, so the only relevant property of a fragment is the number of subsystems it contains. In this case, the label $F_k$ effectively denotes any fragment consisting of $k$ environmental subsystems, and one can only look at the mutual information between the system and varying-sized fragments of the environment. Under these conditions, we say that Quantum Darwinism (QD) appears when fragments $F_k$ with $k$ in a wide range of values roughly contain all the information about the system's state. In this case, the environment contains many redundant copies of the system's state, and one can thus say this became objective. In practice, this behaviour is signalled by the appearance of the so-called redundancy plateau in the function
\begin{equation}
    R(k)=\mathcal{I}(\Gamma:F_k)
    \label{e.R(k)}
\end{equation}
at a value $R(k)\simeq S(\Gamma)$ starting from around the value $k_{\rm min}$ of the minimal fragment size containing enough information to describe a copy of the system's state. This plateau signals that having extra access to the environmental subsystems does not increase the information one has about the system's state. In the literature, it is often said that under these conditions, the state of the system is redundantly encoded in the environment, which implies it could be decoded by an observer having access to a fragment of size larger than $|F_{k_{\rm min}}|$ only. Yet, the precise meaning of this sentence is rarely explored formally. This is what we accomplished in this work.

\subsection{Recovery channels}
\label{s.recovery}

When a quantum system interacts with an environment, it typically evolves to lose quantum coherence. This evolution is most generally described by a quantum channel and is generally irreversible due to information loss: once a system undergoes open dynamics, it is usually impossible to return it to its original state by simple means. (In contrast, only unitary dynamics are perfectly reversible.)  Nevertheless, some open evolutions admit perfect or approximate recovery channels~\cite{Wilde15}. These represent physically implementable processes capable of restoring the state without literally “rewinding” the lossy dynamics, as one would do when the evolution is unitary. Following~\cite{LautenbacherEtAl22}, we review the basic properties of recovery channels and outline the main steps for identifying the optimal recovery method for a given system.

We start by considering the following definition: a completely positive and trace preserving (CPTP) map $\Lambda:\mathcal{D}(A)\rightarrow\mathcal{D}(B)$ is said to be recoverable over $\mathcal{S}(A)\subset \mathcal{D(}A)$ if there exists another CPTP map $\mathcal{R}^\mathcal{S}_\Lambda:\mathcal{D(}B)\rightarrow\mathcal{D(}A)$, such that 
\begin{equation}
    \mathcal{R}^\mathcal{S}_\Lambda[\Lambda(\rho)]=\rho
\end{equation}
for all $\rho\in\mathcal{S}(A)$, which are called the recoverable states. The map $\mathcal{R}^{\mathcal{S}(A)}_\Lambda$ is called the recovery map of $\Lambda$ for the subset $\mathcal{S}(A)$. Given a quantum channel $\Lambda$ and a set $\mathcal{S}(A)$ we want to recover, finding the correct recovery map is not easy. One such construction was found by Petz~\cite{Petz86, Petz03}: once the system in the state $\rho_i$ has evolved through the channel $\Lambda$ to the state $\rho_f=\Lambda(\rho_i)$, we may try to recover the initial state by the Petz recovery map:
\begin{equation}
    \mathcal{P}_{\Lambda}^{\sigma}(\rho_f)\equiv \sigma^{\frac{1}{2}}\Lambda^{\dagger}[\Lambda(\sigma)^{-\frac{1}{2}}\rho_f\Lambda(\sigma)^{-\frac{1}{2}}]\sigma^{\frac{1}{2}}~,
    \label{e.Petz}
\end{equation}
where $\sigma\in\mathcal{D(A)}$ is a reference state, $\Lambda^{\dagger}:\mathcal{L(H_B)}\rightarrow\mathcal{L(H_A)}$ is the trace-dual of $\Lambda$, and all inverses are Moore–Penrose inverses. It is possible to show that the Petz map is well defined on the image of $\Lambda$, at least once $\sigma$ is selected to be full-rank (see App.~\ref{s.SuppPetz}). We characterise the subset of states for which a quantum map is recoverable by the fact that the relative entropy between them and the reference state does not change due to the action of $\Lambda$. In other words, a state $\rho_f$ can be perfectly recovered to its initial value $\rho_i$ by $\mathcal{P}^\sigma_\Lambda$ if
\begin{equation}\label{e.RelEnt}
S(\rho_i||\sigma)=S(\Lambda(\rho_i)||\Lambda(\sigma))~,
\end{equation}
with the relative entropy between two states defined as $S(\rho||\sigma)=\text{Tr}[\rho\text{log}(\rho)-\rho\text{log}(\sigma)]$.  Notice that the RHS of the above is well-defined anywhere the LHS is, as reviewed in App.~\ref{s.SuppPetz}. Once a recovery map is constructed on a subset, it can be extended to larger regions or even to the entire Hilbert space, obtaining a strategy for an approximate recovery. The effectiveness of the resulting approximate recovery must then be assessed on a case-by-case basis. One way to evaluate its performance is using some measure of the quality of the recovery across the system's state space. One such measure is given by the fidelity
\begin{equation}
    F(\rho_i,\rho_r)=\text{Tr}\sqrt{\sqrt{\rho_i}\rho_r\sqrt{\rho_i}}
\end{equation}
between the initial ($\rho_i$) and recovered ($\rho_r$) states~\cite{LautenbacherEtAl22, LautenbacherEtAl24}. The Fawzi–Renner theorem guarantees the existence of a recovery map (which can be chosen as a rotated Petz map) achieving fidelity at least $e^{-\Delta/2}$, with
\begin{equation}
    \Delta=S(\rho||\sigma)-S(\Lambda(\rho)||\Lambda(\sigma))
\end{equation}
\cite{FawziR15}. In the special case of $\Delta=0$, Petz's recovery theorem also implies that the ordinary Petz map \eqref{e.Petz} perfectly recovers the state, yielding unit fidelity~\cite{Petz03}.

\section{Theoretical analysis}

Let us consider a principal system $\Gamma$ interacting with a multipartite environment $\Xi=\cup_{i=1}^N\Xi_i$, with $\Xi_i\cap\Xi_j=\emptyset$ for all $i\neq j$. We group the parts of the environment in a collection of fragments $\{F_k\}$ such that 
\begin{equation}
    F_k=\cup_{i=1}^k\Xi_i~.
\end{equation}
It holds that $F_k\subset F_{k'}$ for all $k< k'$, and $F_N=\Xi$. Moreover, we assume $\Gamma$ and $\Xi$ are coupled via a measurement-like interaction $\hat{U}$, as described in Sec.~\ref{s.QD}. In what follows, we require the interaction to be (approximately) permutation invariant across $\Xi$. With this setup, our aim in this section is to describe the encoding of information from $\Gamma$ to the $F_k$'s under $\hat{U}$, describe if and how the Petz recovery map can be used to reconstruct the state of the former from that of the latter, and characterise how the quality of recovery varies with the fragment size, parametrised by $k$.

For instance, $\Gamma=s_0$ may represent a spin located at the centre of a ring of spins $\Xi=\cup_{i=1}^N s_i$ labelled from $1$ to $N$ in the clockwise direction. In this geometry, one may define the fragments as $F_k=\{s_1,\dots,s_k\}$. In this example, permutation invariance means that the central spin interacts (almost) equally with every spin in the ring. Alternatively, we could place the distinguished spin at one end of a linear array of spins (a spin chain without internal interactions). However, our request for permutation invariance implies that the interaction cannot depend on the distance along the chain, which restricts the allowed couplings to non-physical ones. For this reason, examples with rotationally symmetric (or sufficiently random) geometries are typically preferred in the literature.

Starting from a given separable state of the total system 
\begin{equation}
    \rho_\Psi=x_\Gamma\otimes X_\Xi~,
\end{equation}
the interaction encodes the system's state in the environment's one. Keep an eye on notation: here and hereafter, we use lowercase Roman letters for states of $\Gamma$, uppercase Roman letters for states of (parts of) the environment $\Xi$, and lowercase Greek letters for states of generic systems, of $\Psi$, and of partitions of $\Psi$ spread across both $\Gamma$ and the environment. For each fragment $F_k$, we define the map $\Lambda_{k}:\mathcal{D}(\Gamma)\to\mathcal{D}(F_k)$ as
\begin{equation}
    \Lambda_{k}(\bullet)={\rm Tr}_{\Gamma\cup(\Xi\setminus F_k)}[\hat{U}(~\bullet~\otimes  X_\Xi)\hat{U}^\dagger]~;
    \label{e.encoding_channel}
\end{equation}
these maps describe how the initial state of $\Gamma$ gets encoded in the final state of the fragment $F_k$. From these, we build the associated Petz recovery channel $\mathcal{P}_{\Lambda_k}^s:\mathcal{D}(F_k)\to\mathcal{D}(\Gamma)$, with reference state $s\in\mathcal{D}(\Gamma)$ fixed across all channels $\Lambda_k$.

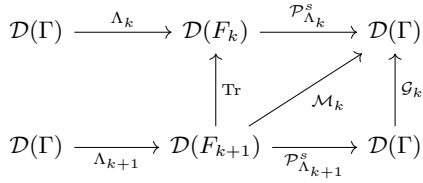
\begin{figure}
  \centering

\begin{tikzcd}[row sep=3em, column sep=3.5em]
  \mathcal{D}(\Gamma)
    \arrow[r, "\Lambda_k"]
    %\arrow[dr, "\Lambda_{k+1}"']
  & \mathcal{D}(F_k)
    \arrow[r, "\mathcal{P}^s_{\Lambda_k}"]
  & \mathcal{D}(\Gamma) \\
  \mathcal{D}(\Gamma) \arrow[r, "\Lambda_{k+1}"']
  &
  \mathcal{D}(F_{k+1})
    \arrow[u, "{\rm Tr}"'] \arrow[ur, "\mathcal{M}_k"']
    \arrow[r, "\mathcal{P}^s_{\Lambda_{k+1}}"']
  & \mathcal{D}(\Gamma) \arrow[u, "\mathcal{G}_k"']
\end{tikzcd}

  \caption{A diagram showing all maps we need in our arguments.}
  \label{fig:comm-diagram}
\end{figure}

Having introduced the necessary formal elements, we can now state our aim more precisely: we study the quality of recovery through the root fidelity
\begin{equation}
    Q(k)=F(x_\Gamma,(\mathcal{P}_{\Lambda_k}^s\circ\Lambda_k)(x_\Gamma))
\end{equation}
seen as a function of $k$. Proofs of all the following statements can be found in App.~\ref{a.M}. We begin by noticing that
\begin{equation}
    \Lambda_k={\rm Tr}_{\Xi_{k+1}}\circ \Lambda_{k+1}~,
\end{equation}
which in turn implies
\begin{equation}
    \Lambda_k^\dagger(X_{F_k})=\Lambda_{k+1}^\dagger(\mathbb{I}_{\Xi_{k+1}}\otimes X_{F_k})~.
\end{equation}
Next, we write
\begin{equation}
    \mathcal{P}_{\Lambda_k}^s\circ \Lambda_k=\mathcal{M}_k\circ\Lambda_{k+1}
\end{equation}
where we define $\mathcal{M}_k:\mathcal{D}(F_{k+1})\rightarrow\mathcal{D}(\Gamma)$ as
\begin{equation}
    \mathcal{M}_k(\bullet)\equiv s^{\frac{1}{2}}\Lambda_{k+1}^\dagger[\mathbb{I}_{\Xi_{k+1}}\otimes(\Lambda_k(s)^{-\frac{1}{2}}{\rm Tr}_{\Xi_{k+1}}[~\bullet~]\Lambda_k(s)^{-\frac{1}{2}})]s^{\frac{1}{2}}~.
    \label{e.M}
\end{equation}
The map $\mathcal{M}_k$ can be understood as an imperfect $(k+1)$-th Petz recovery that, when applied to the larger fragment $F_{k+1}$, removes any possible extra information encoded in $\Xi_{k+1}$ to give the same result as the $k$-th Petz acting on the fragment $F_k$. The map $\mathcal{M}_k$ is TP and has adjoint
\begin{equation}
    \mathcal{M}_k^\dagger(\bullet)=\mathbb{I}_{\Xi_{k+1}}\otimes\Lambda_k(s)^{-\frac{1}{2}}\Lambda_{k}(s^{\frac{1}{2}}~\bullet~ s^{\frac{1}{2}})\Lambda_k(s)^{-\frac{1}{2}}~.
\end{equation}
$\mathcal{M}_k$ can also be written as: 
\begin{equation}
    \mathcal{M}_k=\mathcal{G}_k\circ\mathcal{P}^s_{\Lambda_{k+1}}~,
    \label{e.M_prop_R}
\end{equation}
where $\mathcal{G}_k:\mathcal{D}(\Gamma)\rightarrow \mathcal{D}(\Gamma)$ reads as the ``imperfection''
\begin{equation}
    \mathcal{G}_k(x)=s^{\frac{1}{2}}\Lambda_{k+1}^\dagger(\mathcal{S}_k(x))s^{\frac{1}{2}}
\end{equation}
we need to apply after the $(k+1)$-th Petz recovery channel to obtain the output of the $k$-th one (see Fig.~\ref{fig:comm-diagram}). In the above, $\mathcal{S}_k:\mathcal{D}(\Gamma)\rightarrow\mathcal{D}(F_{k+1})$ is a supplementary map defined by its action on the image of the Petz as prescribed in App.~\ref{a.M}. The map $\mathcal{G}_k$ is TP on ${\rm Im}(\mathcal{P}^s_{\Lambda_{k+1}}\circ\Lambda_{k+1})\subseteq{\rm Im}(\mathcal{P}^s_{\Lambda_{k}})$, but it is not necessarily CP, even if $\mathcal{M}_k$ and $\mathcal{P}^s_{\Lambda_{k+1}}$ are.

We know from its definition that $\mathcal{G}_k$ acts as the identity on all those states that are equally recovered from $F_k$ and $F_{k+1}$, \textit{i.e.}
\begin{equation}
    \lvert\lvert\mathcal{G}_k(x')-x'\rvert\rvert=\lvert\lvert\mathcal{P}^s_{\Lambda_{k}}\circ\Lambda_{k}(x)-\mathcal{P}^s_{\Lambda_{k+1}}\circ\Lambda_{k+1}(x)\rvert\rvert~.
\end{equation}
Therefore, the closer the $k$- and the $(k+1)$-recovered states are, the more $\mathcal{G}_k$ acts as the identity on the latter (and hence that state turns out to be a fixed point of $\mathcal{G}_k$). In particular, if the extra subsystem $\Xi_{k+1}$ carries no extra information about $\Gamma$, then the $(k+1)$-th Petz map already forgets about it, $\mathcal{P}^s_{\Lambda_{k+1}}\circ\Lambda_{k+1}(x)=\mathcal{P}^s_{\Lambda_{k}}\circ\Lambda_{k}(x)$, and $\mathcal{G}_k$ acts trivially on $\mathcal{P}^s_{\Lambda_{k+1}}\circ\Lambda_{k+1}(x)$. Good recovery at increasingly large fragments forces $\mathcal{G}_k$ to act ever more like the identity on the relevant states.

We now wish to clarify what it means when $Q(k)$ reaches a plateau, that is, we want to understand if and how the condition $\Delta Q_k\equiv Q(k+1)-Q(k)\simeq0$  relates to the usual notion of redundancy plateau, characterised by the flattening of \eqref{e.R(k)}. To this end, we consider the conditional mutual information for a tripartite system $\Psi=A\cup B=A\cup(B'\cup B'')$, defined as
\begin{equation}
    \mathcal{I}(A:B''|B')=\mathcal{I}(A:B)-\mathcal{I}(A:B')~,
\end{equation}
and choose $A=\Gamma$, $B'=F_k$, $B''=\Xi_{k+1}$, and hence $B=B'\cup B''=F_{k+1}$. Then, the redundancy plateau, characterised as
\begin{equation}
    \mathcal{I}(\Gamma:F_{k})=\mathcal{I}(\Gamma:F_{k+1})~,
\end{equation}
corresponds to 
\begin{equation}
    \mathcal{I}(\Gamma:\Xi_{k+1}|F_k)=0~.
    \label{e.RP_via_cond_mutual_info}
\end{equation}
This is the condition for $\Gamma$, $F_k$, and $\Xi_{k+1}$ to form a quantum Markov chain, which means that adding $\Xi_{k+1}$ gives no additional information about $\Gamma$ beyond what was already known from having $F_k$. The above condition also implies that there exists a perfect recovery channel such that:
\begin{equation}
    \rho_{\Gamma\cup F_{k+1}}=({\rm id}_\Gamma\otimes\mathcal{R}_{F_k\rightarrow F_{k+1}})(\rho_{\Gamma\cup F_{k}})~;
    \label{e.Petz_D_proof}
\end{equation}
the above is obtained by taking $\rho=\rho_{\Gamma\cup F_{k+1}}$, $\sigma=\rho_{\Gamma}\otimes\rho_{F_{k+1}}$, and $\Lambda={\rm Tr}_{\Xi_{k+1}}$ in the Petz theorem (see App.~\ref{a.Gamma from F_k}). In practice, the fact that \eqref{e.Petz_D_proof} is a perfect recovery map implies that if we have a protocol $\mathcal{T}_{F_{k+1}\rightarrow\Gamma}$ to recover $\Gamma$ from $F_{k+1}$, we can also construct an equally good strategy to recover $\Gamma$ from $F_k$ as:
\begin{equation}
    \mathcal{T}_{F_{k}\rightarrow\Gamma}=\mathcal{T}_{F_{k+1}\rightarrow\Gamma}\circ\mathcal{R}_{F_k\rightarrow F_{k+1}}~,
\end{equation}
meaning that the initial state of $\Gamma$ can be perfectly reconstructed by acting on $F_k$ alone~\cite{HaydenEtAl04}. In summary, because adding $\Xi_{k+1}$ does not add any information about $\Gamma$ to what we already have from $F_k$, and, in this case, the Petz theorem guarantees that the marginal state for $\Gamma$ obtained from $\rho_{\Gamma F_{k+1}}$ and $\rho_{\Gamma F_k}$ are the same, and hence a plateau of $Q(k)$ always appears in correspondence of a plateau of $R(k)$.

The converse is not true: as $\Delta Q_k=0$ only signals a constant fidelity between the initial state and the recovered state, plateaus of $Q(k)$ do not automatically imply a redundancy of encoding. Several recovered states may have the same fidelity with the initial state, yet differ and encode distinct information about $\Gamma$.

\subsection{Recovery of the fittest} 
In this section, we use a simple but important example to show if and how the above Petz maps $\mathcal{P}^s_{\Lambda_k}$ can be used to reconstruct the state of the principal system from the environmental fragments $F_k$. Specifically, we select the interaction to be an Ozawa-like premeasurement, as presented in Sec.~\ref{s.QD}. The same interaction also models strong environment-induced decoherence onto an einselected basis, a scenario known to display quantum Darwinism behaviour~\cite{BlumeKohoutZ05}. For our model, we take a one-to-all qubit system, where the one acts as the principal system and the all as the measurement apparatus/environment~\cite{Zurek03, PranziniV24}. While not necessary, we take the all to start in the ready-to-read\footnote{A ready-to-read state is any pure state of the environment that is not an eigenstate of the environmental component $O_\Xi$ of the system–environment interaction \eqref{e.H_generic}.} state:
\begin{equation}
    \ket{\Xi^R}=\bigotimes_{l=1}^N\ket{\Xi^R}_l%\equiv\bigotimes_{l=1}^N\left(\alpha_l\ket{0}_l+\beta_l\ket{1}_l\right)~,
    \label{e.ready_state}
\end{equation}
and the one in the initial state:
\begin{equation}
    x_\Gamma=\sum_{r,r'=0,1}\gamma_{rr'}\ket{r}\bra{r'}~.
\end{equation}
The basis $\{\ket{r}~s.t.~r=0,1\}$ we used above is the one diagonalising the system's part of the interaction Hamiltonian 
\begin{equation}
H=O_\Gamma\otimes\sum_{l=1}^N g_l~ \mathbb{I}_2^{\otimes(l-1)}\otimes O_l\otimes\mathbb{I}_2^{\otimes(N-l)}~,
\label{e.H}
\end{equation}
with
\begin{equation}
    O_\Gamma=\sum_r\omega_r\ket{r}\bra{r}~,
\end{equation}
where the specific form of the operators $O_l$ is left free at this stage. The dynamics couples the system and the apparatus, and after some time $t$, it gives the joint state
\begin{equation}
    \rho(t)=\sum_{r,r'}\gamma_{rr'}\ket{r}\bra{r'}\otimes\ket{\Xi^r(t)}\bra{\Xi^{r'}(t)}~,
\end{equation}
where 
\begin{equation}
    \ket{\Xi^r (t)}=\bigotimes_{l=1}^N U^r_l(t)\ket{\Xi^R}_l=\bigotimes_{l=1}^N\ket{\Xi^r (t)}_l
\end{equation}
with 
\begin{equation}
    U^r_l(t)=\exp[-i\omega_r g_lO_l t]~.
\end{equation}
So far, this is only an implementation of the general framework reviewed in Sec.~\ref{s.QD}. From this, it is easy to obtain the encoding channel \eqref{e.encoding_channel} as:
\begin{equation}
    \Lambda_k(t;x_\Gamma)=\sum_r\gamma_{rr}\bigotimes_{l=1}^k\ket{\Xi^r(t)}_l\bra{\Xi^{r}(t)}_l
\end{equation}
As we expected, any information about the off-diagonal elements of the principal system's state is lost in the encoding. The Kraus operators, introduced by
\begin{equation}
    \Lambda_k(t;x_\Gamma)=\sum_{r} E^{(r)}_k(t)x_\Gamma E^{(r)\dagger}_k(t)
\end{equation}
are easily computed as:
\begin{equation}
    E^{(r)}_k(t)=\bigotimes_{l=1}^k\ket{\Xi^r(t)}_l\bra{r}
\end{equation}
and of course satisfy $\sum_rE^{(r)\dagger}_k(t) E^{(r)}_k(t)=\sum_r\ket{r}\bra{r}=\mathbb{I}_\Gamma$. We now have all the elements we need to construct the Petz recovery map. First, the Kraus operators give the adjoint channel
\begin{equation}
    \Lambda^\dagger_k(t;X_{F_k})=\sum_{r} E^{(r)\dagger}_k(t) X_{F_k}E^{(r)}_k(t)~.
\end{equation}
Next, we choose the reference state $\sigma=\mathbb{I}_\Gamma/2$ for the sake of simplicity, and obtain
\begin{equation}
\begin{split}
    \Lambda_k(t;\sigma)&=\frac{1}{2}\sum_r\bigotimes_{l=1}^k\ket{\Xi^r(t)}_l\bra{\Xi^{r}(t)}_l\\
    &=\frac{1}{2}\sum_r E^{(r)}_k(t)E^{(r)\dagger}_k(t)~.
\end{split}
\end{equation}
Taking the inverse square root of $\Lambda_k(\sigma)$ is not trivial, unless it reduces to a sum of orthonormal projections. This only happens at those times when the probabilities associated with projectors onto einselected states of the system are transferred to probabilities associated with projectors onto orthogonal states of the environment. These times are said to satisfy the probability reproducibility condition (PRC), and denoted by $t_{{\rm PRC}}$~\cite{PranziniV24, HeinosaariZ11}. At PRC times we get
\begin{equation}
    x_\mathcal{R}=\mathcal{P}^\sigma_{\Lambda_k}(t_{{\rm PRC}};X_{F_k})=\sum_rp_r\ket{r}\bra{r}
\end{equation}
with 
\begin{equation}
    p_r=\braket{\Xi^r(t_{{\rm PRC}})|X_{F_k}|\Xi^r(t_{{\rm PRC}})}=\gamma_{rr}
\end{equation}
where we used that $X_{F_k}=\Lambda_k(x_\Gamma)$ and the fact that $p_r$ is easy to calculate at the PRC times. Computing the fidelity between $x_\Gamma$ and $x_\mathcal{R}$ gives
\begin{equation}
    F(x_\Gamma,x_\mathcal{R})^2=1-2\left(p_0p_1-\sqrt{p_0p_1(p_0p_1-|\gamma_{01}|^2)}\right)~.
    \label{e.F_at_PRC}
\end{equation}

This simple result reveals two things. First, unit fidelity can be achieved iff $|\gamma_{01}|=0$, that is, if the initial state was a convex combination of pointer states. This means only ``classically objective'' mixtures undergo the environmental encoding in such a way that they fully survive and can thus be completely decoded from the environmental fragments. Yet, note that even in this case, unit fidelity can only be achieved at PRC times, and one should expect non-maximal fidelity at all other times.

Second, we observe that the above (PRC) fidelity does not depend on $k$, indicating that fragments of any size suffice for state recovery. Once again, this is an artefact of considering PRC times and of our simple model displaying perfect permutation invariance. In fact, at PRC times, all parts of the environment redundantly encode the same information, which is also enough for pointer state reconstruction. This does not hold in general, but obtaining a non-trivial dependence on $k$ requires taking powers of $\Lambda_k(t;\sigma)$ at generic times and using them to compute the Petz recovery map, a task which can be formidable if approached analytically. The next section addresses this point via numerical methods, as well as the fact that $Q(k)$ still depends on $k$ even at PRC times if permutation invariance is relaxed.

\begin{figure}
    \centering
    \includegraphics[width=1.05\linewidth]{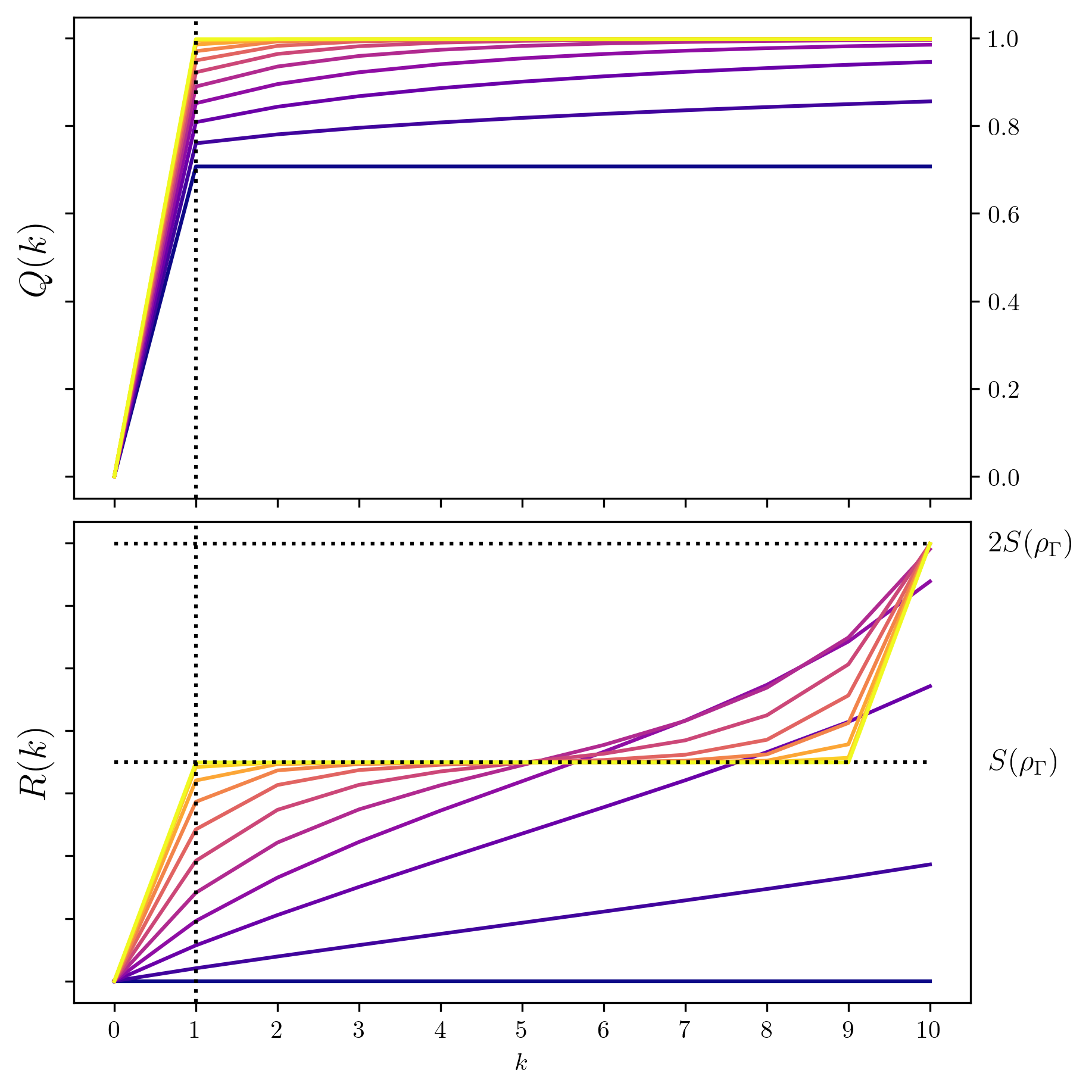}
    \caption{$Z$-$Z$ coupling with fixed $g_k=g=1$, pure initial state of $\Gamma$ close to a pointer state ($r=1$, $\theta=0.1$, and $\phi=0$ on the Bloch sphere) and environment initial state $\ket{\Xi^R}=\ket{+}^{\otimes N}$, with $N=10$. Quality of Petz recovery (above) and mutual information (below) are plotted as functions of the fragment size, and for ten evenly spaced interaction times between $t_i=0$ and $t_f=\pi/(4g)$ (darker is earlier). Note that $t_f$ is one of the PRC times $t_{\rm PRC}=\pi(1+2n)/4g$, $n\in\mathbb{Z}$, which are periodic in this case. The horizontal dotted lines in the plot below are the von Neumann entropy $S(\rho_\Gamma)$ and $2S(\rho_\Gamma)$ for $\rho_\Gamma=\rho_\Gamma(t_f)$ (yellow curve), and the vertical dotted line is the first value of $k$ for which $R(k)\geq S(\rho_\Gamma)$.}
    \label{fig:ZZ-ordered-pure-pointer}
\end{figure}

\section{Numerical analysis}
We now show both the above results and those points we could not address analytically by numerical analysis. For concreteness, we specialise the above one-to-all measurement-like interaction to:
\begin{enumerate}
    \item[\textbf{A}] a collection of $N$ identical $Z$-$Z$ interactions with identical couplings, obtained by selecting $O_\Gamma=Z$ and $O_l=Z$ in Eq.~\eqref{e.H};
    \item[\textbf{B}] a collection of $N$ random $Z$-$H$ interactions, obtained by selecting $O_\Gamma=Z$ and $O_l=H_l$ random hermitian matrices drawn from the ${\rm GUE}(2)$ ensemble in Eq.~\eqref{e.H}.
\end{enumerate}
The cases \textbf{A} and \textbf{B} differ in one main respect. The first is fully permutation invariant, so the emergence of plateaus in $R(k)$ and $Q(k)$ is entirely expected. The second case, by contrast, is not strictly permutation invariant, since all couplings differ due to the random choice of $O_l$. Nevertheless, we expect this microscopic randomness to become irrelevant when considering large enough fragments, effectively restoring an effective notion of permutation invariance of, and thus recoverability from, sufficiently large fragments. Consequently, we expect the qualitative behaviour observed in the trivial, fully symmetric case \textbf{A} to carry over to the non-permutation invariant case \textbf{B}, albeit only once fragments are large enough for the randomness to average out.

Additionally, one can interpolate between the two cases above by choosing the couplings $g_l$ at random in the $Z$-$Z$ interactions. This effectively rescales the eigenvalues of the environmental $Z$ operators, making them random but not distributed as ${\rm GUE}(2)$. Since this variant does not change the qualitative behaviour of the recovery, we do not discuss it further.

\begin{figure*}
    \centering
    \begin{subfigure}{0.46\textwidth}
        \includegraphics[width=1.1\linewidth]{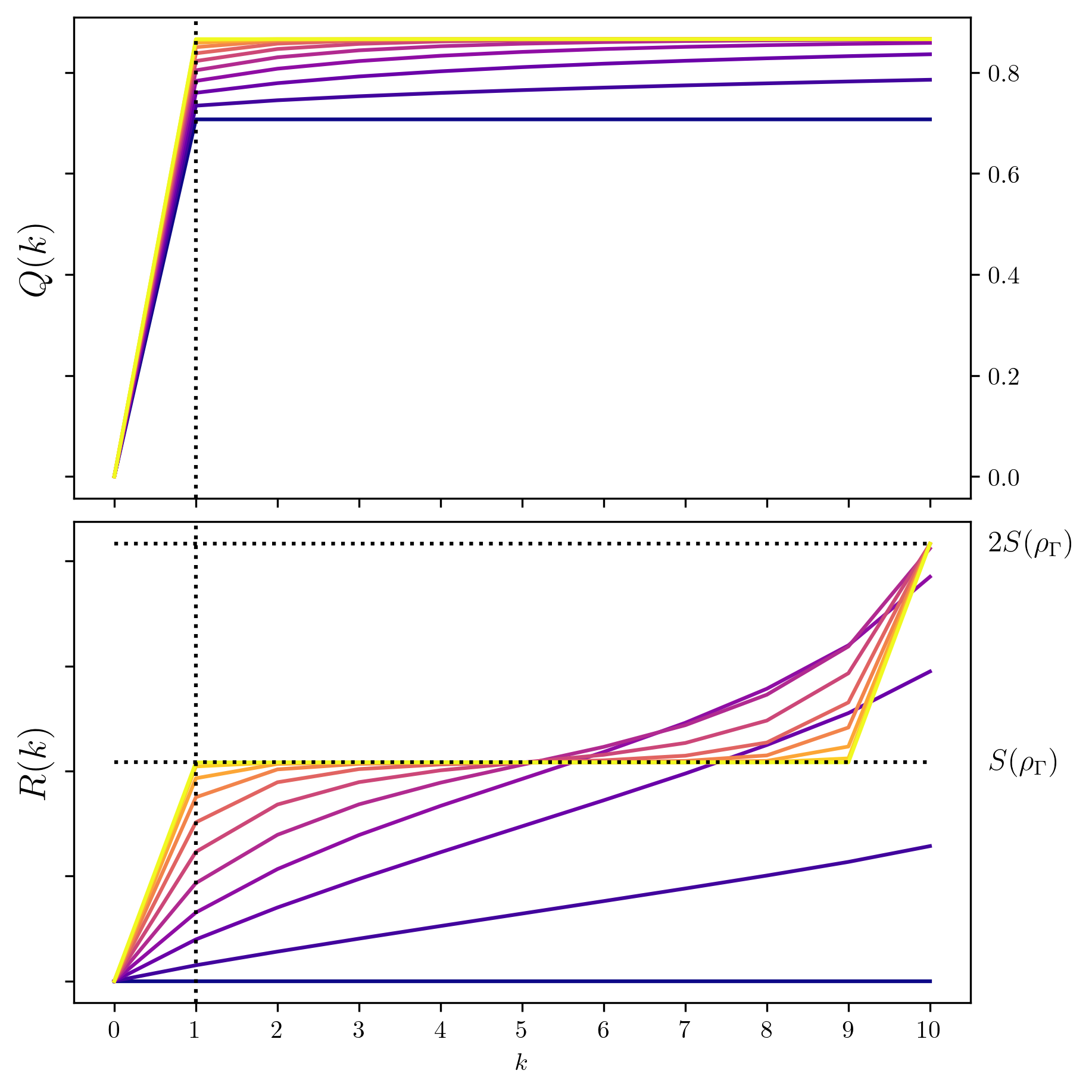}
    \caption{$Z$-$Z$ coupling with initial state of gamma pure and far from any pointer state ($r=1$, $\theta=\pi/4$, and $\phi=0$ on the Bloch sphere). All other details as in Fig.~\ref{fig:ZZ-ordered-pure-pointer}.}
    \label{fig:ZZ-ordered-pure-not-pointer}
    \end{subfigure}
    \hspace{2em}
    \begin{subfigure}{0.45\textwidth}
        \includegraphics[width=1.1\linewidth]{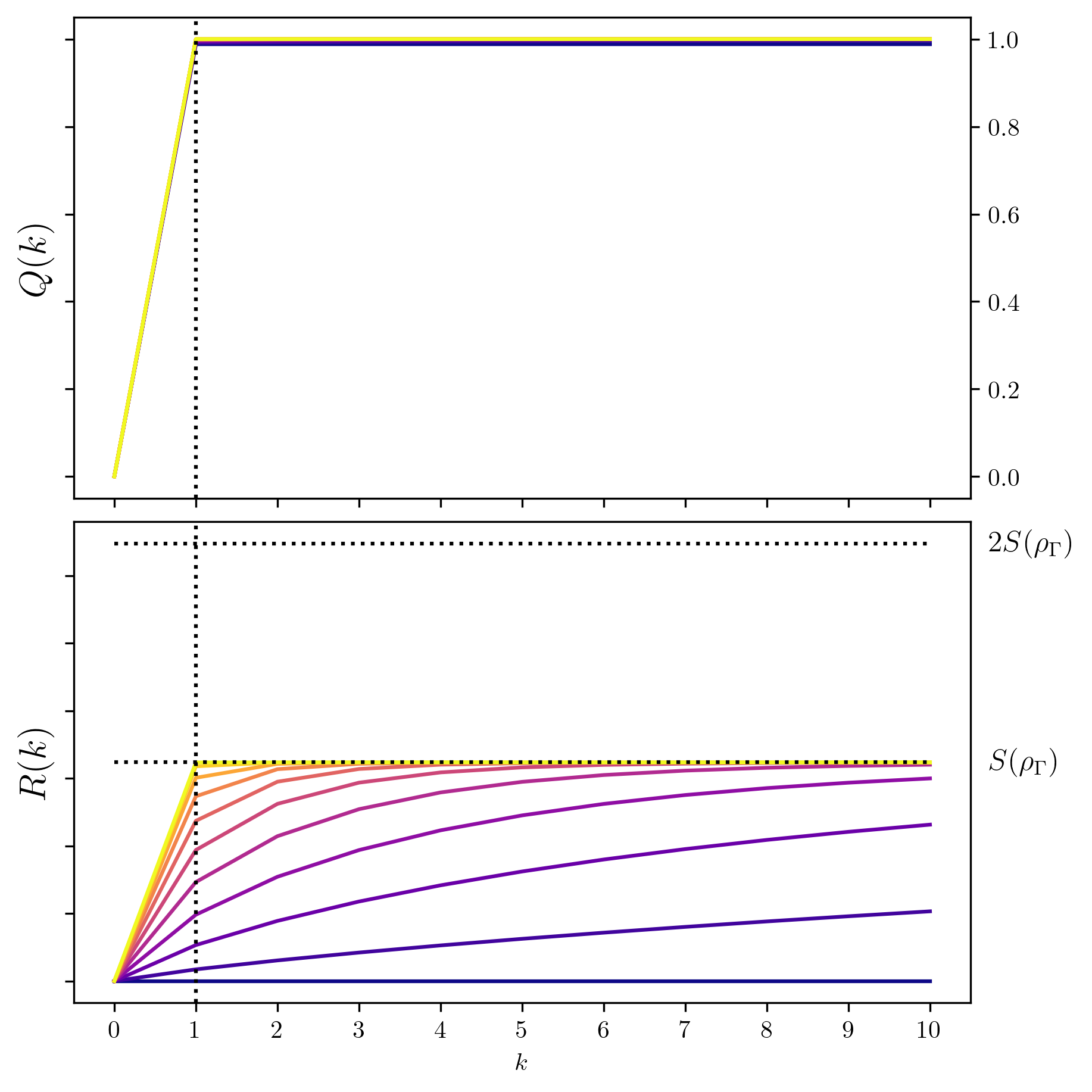}
    \caption{$Z$-$Z$ coupling with initial state of gamma a mixture of pointer states ($r=0.3$, $\theta=0$, and $\phi=0$ on the Bloch sphere). All other details as in Fig.~\ref{fig:ZZ-ordered-pure-pointer}.}
    \label{fig:ZZ-ordered-mixed-pointer}
    \end{subfigure}
    \caption{$Z$-$Z$ coupling with fixed $g_k=g=0.1$ for a pure non-pointer state (left) and mixture of pointer states (right).}
\end{figure*}

For our numerical simulations, we select the initial state of the system to either be:
\begin{enumerate}
    \item[\textbf{1}] a pure state close to a pointer state;
    \item[\textbf{2}] a convex combination of pointer states;
    \item[\textbf{3}] a pure state far away from any pointer state.
\end{enumerate}
The numerical analysis is performed using QuTiP~\cite{qutip5}. 

The case \textbf{A1} is shown in Fig.~\ref{fig:ZZ-ordered-pure-pointer}, with $g_k=g$ for all $k$, pure initial state of $\Gamma$ close to a pointer state and environment initial state. This is the typical setting for which we expect the redundancy plateau to appear, as is indeed shown in the figure (bottom). The figure also shows (top) that the quality of recovery plateaus as soon as the environment is large enough to encode some information about the system, but it caps at one only when the redundancy plateau is realised in the mutual information. That is, almost-constant fidelity of recovery is always achieved, but the quality is sub-optimal for all times not displaying the redundancy plateau. 

In the case \textbf{A2}, shown in Fig.~\ref{fig:ZZ-ordered-pure-not-pointer} with all details as above but the initial state of $\Gamma$ chosen far from any pointer state, we still get a redundancy plateau but the fidelity caps at values lower than one, signalling that the state was indeed not a pointer one. Finally, the case \textbf{A3} is displayed by Fig.~\ref{fig:ZZ-ordered-mixed-pointer}, where we took $\rho_\Gamma(0)=0.8\ket{0}\bra{0}+0.2\ket{1}\bra{1}$.

\begin{figure}
    \centering
    \includegraphics[width=1.05\linewidth]{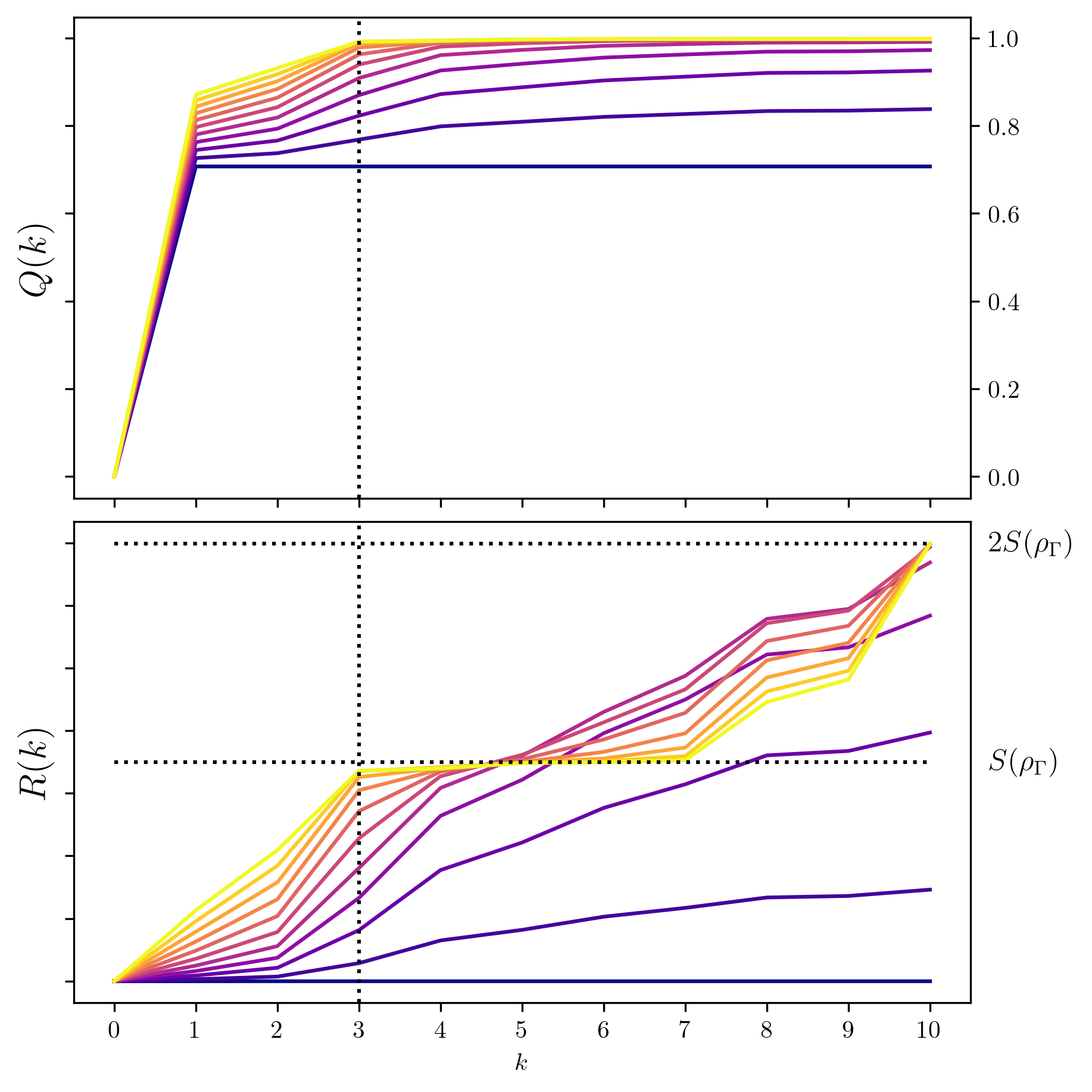}
    \caption{$Z$-$H$ coupling with fixed $g_k=g=1$, and $O_k$ randomly sampled from the ${\rm GUE}(2)$ ensemble as described in the text. All other details as in Fig.~\ref{fig:ZZ-ordered-pure-pointer}.}
    \label{fig:ZH-ordered-pure-pointer}
\end{figure}

\begin{figure*}
    \centering
    \begin{subfigure}{0.46\textwidth}
        \includegraphics[width=1.1\linewidth]{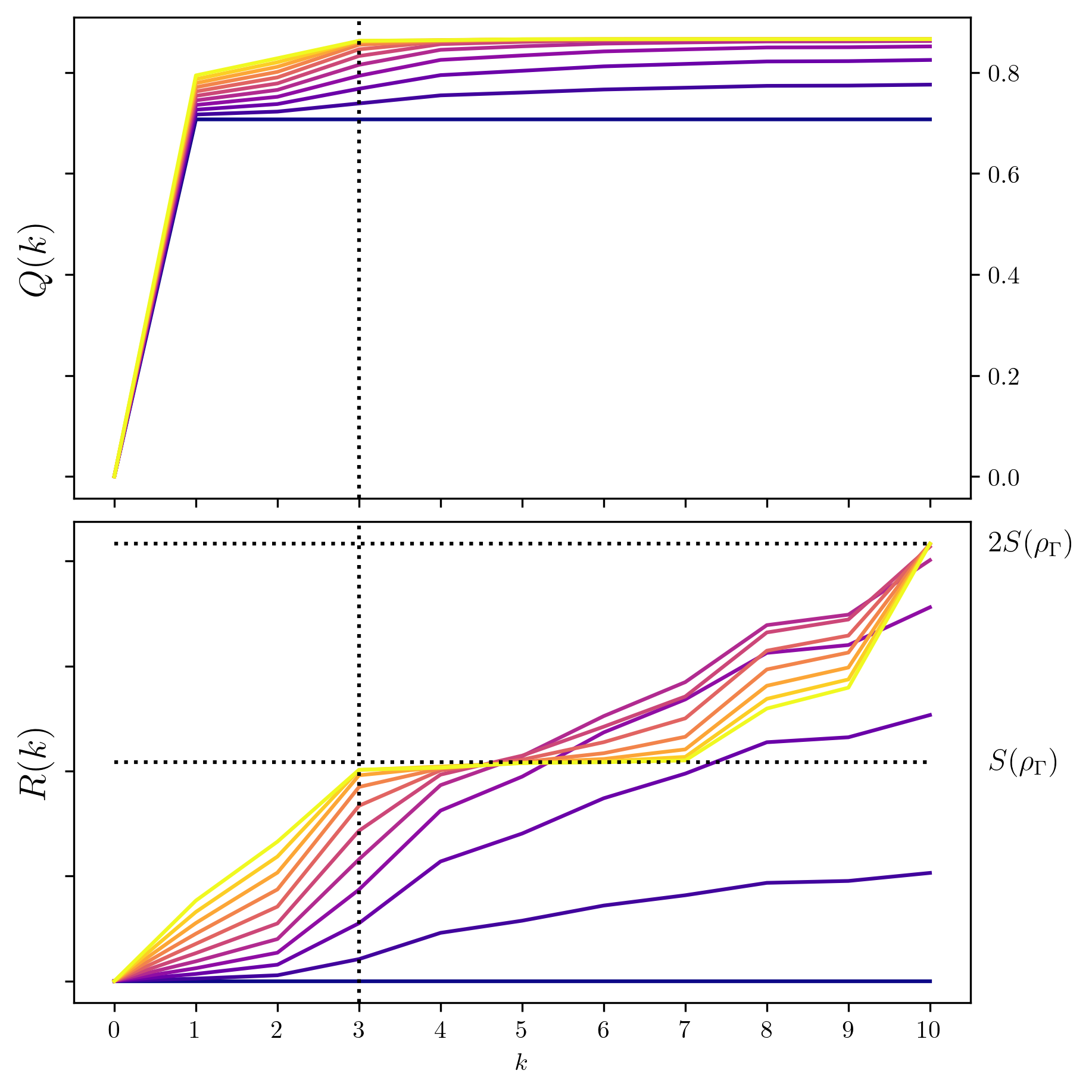}
    \caption{$Z$-$H$ coupling as in Fig. with initial state of gamma pure and far from any pointer state ($r=1$, $\theta=\pi/4$, and $\phi=0$ on the Bloch sphere). All other details as in Fig.~\ref{fig:ZH-ordered-pure-pointer}.}
    \label{fig:ZH-ordered-pure-not-pointer}
    \end{subfigure}
    \hspace{2em}
    \begin{subfigure}{0.46\textwidth}
        \includegraphics[width=1.1\linewidth]{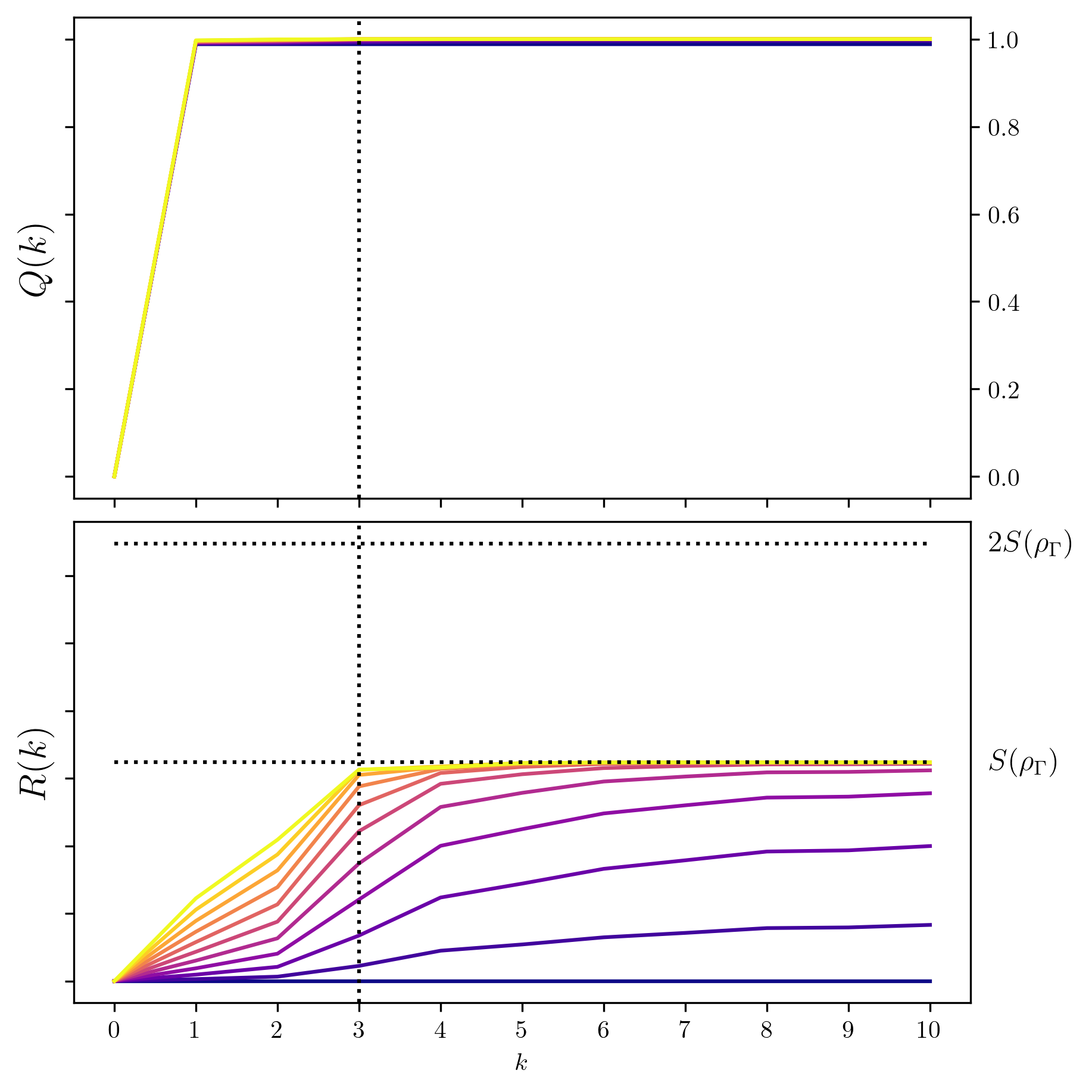}
    \caption{$Z$-$H$ coupling with initial state of gamma a mixture of pointer states ($r=0.3$, $\theta=0$, and $\phi=0$ on the Bloch sphere). All other details as in Fig.~\ref{fig:ZH-ordered-pure-pointer}.}
    \label{fig:ZH-ordered-mixed-pointer}
    \end{subfigure}
    \caption{$Z$-$H$ coupling with fixed $g_k=g=0.1$ for a pure non-pointer state (left) and mixture of pointer states (right).}
\end{figure*}

Let us now study the \textbf{B} cases. For concreteness, we sample the operators $O_k$ from a $\mathrm{GUE}(2)$ ensemble, \textit{i.e.} the Gaussian Unitary Ensemble of $2 \times 2$ Hermitian matrices\footnote{These are random hermitian matrices whose elements are Gaussian distributed (real Gaussian for the diagonal elements, and complex Gaussian for the off-diagonal elements), and such that they are unitary invariant, \textit{i.e.}, $H=UHU^\dagger$ for all $U\in U(n)$. In practice, these are constructed by the normal distribution $\mathcal{N}(0,1)$ as: $H_{ij}=(x_{ij}+iy_{ij})/\sqrt{2}$ with $x_{ij},y_{ij}\sim\mathcal{N}(0,1)$, for $i<j$; $H_{ij}=H_{ji}^*$, for $i>j$; and, $H_{ii}\sim\mathcal{N}(0,1)$ for the diagonal~\cite{AndersonEtAl09}.}. The case \textbf{B1}, displayed in Fig.~\ref{fig:ZH-ordered-pure-pointer}, shows that no redundancy plateau appears in $R(k)$, yet a plateau in $Q(k)$ develops and thus the redundant encoding of information occurs. This generically happens at a larger fragment size. Of course, the result is not in contradiction with the second conclusion following Eq.~\eqref{e.F_at_PRC}, and in fact expected once permutation invariance is broken: put simply, more environmental bits are needed to correct for the errors in the encoding due to the randomness of the interaction. The case \textbf{B2} does not show any interesting new features and is included in Fig.~\ref{fig:ZH-ordered-pure-not-pointer} just for completeness.

%In general, the quality of recovery can be higher if the environment is not symmetric: this can be dramatically seen in the case of an initial state of the environment that is pure but not pointer, as displayed by the case \textbf{B2} in Fig.~\ref{fig:ZH-ordered-pure-not-pointer}. While the plateau in $Q(k)$ is reached at a larger fragment size due to disordered encoding, this is generally placed at a higher value (better recovery) than in the $Z$-$Z$ case, thanks to the fact that random interactions provide the environment with more phase information than the ordered ones. 

Finally, the case \textbf{B3} of a mixture of pointer states is different: a plateau in $R(k)$ is reached at some (generically larger) fragment size, but the quality of fidelity does not reflect this, for it jumps to its maximum value much earlier than the mutual information reaches the plateau. This happens because the environment fragments of any size can become classically correlated with the system, and the Petz recovery no longer benefits from access to larger fragments. Mutual information, however, continues to grow due to random one-to-all couplings that keep generating generic, unstructured entanglement as $k$ increases, which counts as correlation but is irrelevant for recovery.

\section{Conclusions} 
We studied the quality of Petz recovery in system-environment scenarios displaying quantum Darwinism (QD) behaviour. The main objective of our study is to concretise a statement often found in QD literature. This asserts that once the state of a system is redundantly encoded in many fragments of an environment, the former can be reconstructed from the latter. To this end, we constructed a collection of fragment-to-system maps which, starting from the information contained in the environment about the system, try to reconstruct the quantum state of the system as it was before the interaction took place. Then, we showed that the quality of Petz recovery displays a plateau where the mutual information does (but the vice versa is not necessarily true.) 

We approached the problem from both the theoretical and the numerical side. The latter provided us with several pieces of evidence, showing our analysis can apply beyond the case where the principal system is in a pointer state. A large portion of our findings was expected, yet the ability to use a recovery map to \textit{actually} reconstruct the decohered state from the environmental fragments was not, to our knowledge, explicitly shown in the literature. 

Our work is primarily a refinement and consolidation of earlier arguments, but it suggests a possible route for characterizing quantum‑Darwinism–like behaviour in situations where a redundancy plateau in the mutual information is not expected. Since a fidelity plateau is necessary but not sufficient for the classical plateau to arise, it may be that Petz recovery of einselected states from the environment remains feasible (and thus a notion of objectivity still emerges) even in scenarios where the usual redundancy signature via mutual information is absent. A natural setting where this could occur is in constrained systems, where the entropies of the two parts of a bipartite pure state need not coincide, and additional Shannon‑type contributions associated with charges or edge modes can enter the mutual information~\cite{CasiniEtAl14, DonnellyW15, Radicevic14}. These terms may invalidate the standard QD argument, while leaving our recovery strategy intact. 

Finally, we note some interesting parallels between Quantum Darwinism and bulk reconstruction in AdS/CFT, especially when seen as error correction~\cite{CotlerEtAl19, ChenEtAl20, AlmheiriEtAl15}. The emergence of bulk locality in AdS/CFT \cite{Harlow18} and the emergence of classical objectivity in quantum Darwinism share a common information‑theoretic mechanism: both rely on the redundant encoding of a restricted set of observables across many subsystems. In quantum Darwinism, this redundancy underlies objectivity of classical variables, while in AdS/CFT it underlies robustness and locality of bulk observables within a code subspace. While there are many key differences in the details of these two settings, the shared abstract features hint that insights from Quantum Darwinism could be useful in the context of the holographic duality.

\section{Code availability}

To ensure reproducibility, all code used in this study is openly available at \cite{TorvinenEtAl26}. The repository includes: (i) core implementations of the models and recovery maps; and (ii) Jupyter notebooks that reproduce all figures in the manuscript. Details for reproducing the numerical results are provided in the repository. Moreover, additional documentation is included to encourage deeper understanding. 

\section{Acknowledgements}
The authors thank M. Cattaneo and O. Veltheim for valuable comments and discussions, and H. Kwon and M.S. Kim for comments on the first version of the manuscript. JT thanks the Finnish IT Center for Science (CSC), especially the Quantum Technologies group, for valuable support on subject matter topics. NP acknowledges financial support from the Magnus Ehrnrooth Foundation and the Academy of Finland via the Centre of Excellence program (Project No. 336810 and Project No. 336814). NP and EKV acknowledge the financial support of the Research Council of Finland through the Finnish Quantum Flagship project (358878, UH). EKV is in part supported by the Research Council of Finland grant 1371600.

\newpage

\bibliography{bib}

\newpage
\onecolumngrid
\appendix
\section{Support inclusion requirements of relative entropy and their consequences for the Petz map}
\label{s.SuppPetz}
In this section, we briefly review why the quantum relative entropy requires support inclusion and the consequences on the Petz map. The quantum relative entropy appearing in \eqref{e.RelEnt} is only finite if ${\rm Supp}(\rho) \subseteq {\rm Supp}(\sigma)$. More generally, we introduce:
\begin{equation}
     S(\rho || \sigma)=\begin{dcases}
         {\rm Tr}[\rho(\log \rho - \log \sigma)]~~&{\rm if} ~~{\rm Supp}(\rho) \subseteq {\rm Supp}(\sigma)\\
         +\infty ~~&{\rm otherwise}
     \end{dcases}~.
\end{equation}
The reason for this is that, if $\sigma$ has a zero eigenvalue in a direction where $\rho$ has non-zero weight, then $\log \sigma$ has eigenvalue $-\infty$ there, and the trace diverges. However, quantum relative entropy quantifies distinguishability of quantum states via the probability of type II errors $P_{II}\sim\exp(-N S(\rho||\sigma))$ in a hypothesis testing problem with a sample of size $N$, and because if $\rho$ has support outside of $\sigma$ it can be perfectly distinguished from it, then it makes sense to have $S(\rho||\sigma)=+\infty$ in that case. Yet, the above case-by-case approach is cumbersome, and it is easier to avoid the problem by selecting a full-rank $\sigma$ from the beginning. This is always possible when dealing with the Petz map, for we are free to choose our reference at will.

Yet, in many of our arguments and in the Petz map, we used the (Moore-Penrose) inverse of $\Lambda(\sigma)$ and multiplied it by $\Lambda(\rho)$; we need to check if this is sensible. For any completely positive trace-preserving (CPTP) map $\Lambda$, it holds that ${\rm Supp}(\rho) \subseteq {\rm Supp}(\sigma)$ implies ${\rm Supp}(\Lambda(\rho))\subseteq{\rm Supp}(\Lambda(\sigma))$: that is, support inclusion is preserved under CPTP maps. This is readily shown by first noticing that 
\begin{equation}
    {\rm Supp}(\rho) \subseteq {\rm Supp}(\sigma)~~\Leftrightarrow~~\rho\leq c\sigma.
    \label{e.support_inclusion}
\end{equation}
$\Rightarrow$ follows from the fact that $\sigma$ is positive and invertible on its support, and thus $\sigma^{-1/2}$ is well defined there. Then we can consider $\sigma^{-1/2}\rho\sigma^{-1/2}$, which is bounded by some number $c$ times $\mathbb{I}_{{\rm Supp}(\sigma)}$. Thus $\rho\leq c\sigma$. $\Leftarrow$ is also true: given that $\rho\leq c\sigma$ holds, then $\rho$ cannot have any non-zero element in the subspace orthogonal to $\sigma$. Using Eq.~\eqref{e.support_inclusion} and the positivity and linearity of $\Lambda$, we get
\begin{equation}
    c\sigma-\rho\geq 0 ~~\Rightarrow~~\Lambda(c\sigma-\rho)\geq 0~~\Rightarrow~~\Lambda(\rho)\leq c\Lambda(\sigma)~,
\end{equation}
which implies ${\rm Supp}(\Lambda(\rho))\subseteq{\rm Supp}(\Lambda(\sigma))$.

The Petz recovery map involves multiplying its argument by $\Lambda(\sigma)^{-1/2}$, which is well-defined on the support of $\Lambda(\sigma)$. Yet, in the interesting cases, the argument is always the image of the channel $\Lambda$, which the above result guarantees to have support included in the support of $\Lambda(\sigma)$ (at least once $\sigma$ is full-rank).

\section{The imperfect Petz map $\mathcal{M}_k$ and $\mathcal{G}_k$}
\label{a.M}
We begin by explicitly showing how $\mathcal{M}_k$ can be obtained. The composition
\begin{equation}
    \Lambda_k={\rm Tr}_{\Xi_{k+1}}\circ \Lambda_{k+1}~,
\end{equation}
implies
\begin{equation}
    \Lambda_k^\dagger(X_{F_k})=\Lambda_{k+1}^\dagger(\mathbb{I}_{\Xi_{k+1}}\otimes X_{F_k})~.
\end{equation}
due to the definition of adjoint of a map acting on different Hilbert spaces. In fact, considering $\Phi:\mathcal{D}(in)\rightarrow \mathcal{D}(out)$ and an Hermitian operator $X$, we get
\begin{equation}
    (X,\Phi(Y))_{out}={\rm Tr}_{out}[X\Phi(Y)]={\rm Tr}_{in}[\Phi^\dagger(X)Y]=(\Phi^\dagger(X), Y)_{in}~,
\end{equation}
and thus
\begin{equation}
\begin{split}
    (X_{F_k},\Lambda_k(\rho))_{F_k}&={\rm Tr}_{F_k}[X_{F_k}\Lambda_{k}(\rho)]={\rm Tr}_{F_k}[X_{F_k}{\rm Tr}_{\Xi_{k+1}}[\Lambda_{k+1}(\rho)]]={\rm Tr}_{\Xi_{k+1}\cup F_k}[(\mathbb{I}_{\Xi_{k+1}}\otimes X_{F_k})\Lambda_{k+1}(\rho)]\\
    &={\rm Tr}_{\Gamma}[\Lambda_{k+1}^\dagger(\mathbb{I}_{\Xi_{k+1}}\otimes X_{F_k})\rho]=(\Lambda_{k+1}^\dagger(\mathbb{I}_{\Xi_{k+1}}\otimes X_{F_k}),\rho)_\Gamma~.
\end{split}
\end{equation}
Next, we write:
\begin{equation}
    \mathcal{P}_{\Lambda_k}^s(\Lambda_k(x))=(\mathcal{M}_k\circ\Lambda_{k+1})(x)
\end{equation}
where we defined
\begin{equation}
    \mathcal{M}_k(X_{F_{k+1}})\equiv s^{\frac{1}{2}}\Lambda_{k+1}^\dagger(\mathbb{I}_{\Xi_{k+1}}\otimes\Lambda_k(s)^{-\frac{1}{2}}{\rm Tr}_{\Xi_{k+1}}(X_{F_{k+1}})\Lambda_k(s)^{-\frac{1}{2}})s^{\frac{1}{2}}~.
\end{equation}
This identity follows from the previous statements as:
\begin{equation}
        \begin{split}
        \mathcal{P}_{\Lambda_k}^s(\Lambda_k(x))&=s^{\frac{1}{2}}\Lambda_k^{\dagger}(\Lambda_k(s)^{-\frac{1}{2}}\Lambda_k(x)\Lambda_k(s)^{-\frac{1}{2}})s^{\frac{1}{2}}\\
        &=s^{\frac{1}{2}}\Lambda_{k+1}^\dagger(\mathbb{I}_{\Xi_{k+1}}\otimes\Lambda_k(s)^{-\frac{1}{2}}\Lambda_k(x)\Lambda_k(s)^{-\frac{1}{2}}]s^{\frac{1}{2}}\\
        &=s^{\frac{1}{2}}\Lambda_{k+1}^\dagger(\mathbb{I}_{\Xi_{k+1}}\otimes\Lambda_k(s)^{-\frac{1}{2}}{\rm Tr}_{\Xi_{k+1}}[\Lambda_{k+1}(x)]\Lambda_k(s)^{-\frac{1}{2}}]s^{\frac{1}{2}}\\
        &=(\mathcal{M}_k\circ\Lambda_{k+1})(x)
\end{split}
\end{equation}
\subsection{Properties of $\mathcal{M}_k$}
\label{a.properties_M}
The fact that $\mathcal{M}_k$ is TP is found via
\begin{equation}
\begin{split}
    {\rm Tr}_\Gamma[\mathcal{M}_k(X_{F_{k+1}})]&={\rm Tr}_{F_{k+1}}[\mathbb{I}_{\Xi_{k+1}}\otimes\Lambda_k(s)^{-\frac{1}{2}}{\rm Tr}_{\Xi_{k+1}}[X_{F_{k+1}}]\Lambda_k(s)^{-\frac{1}{2}}\Lambda_{k+1}(s)]\\
    &={\rm Tr}_{F_{k}}[\Lambda_k(s)^{-\frac{1}{2}}{\rm Tr}_{\Xi_{k+1}}[X_{F_{k+1}}]\Lambda_k(s)^{-\frac{1}{2}}{\rm Tr}_{\Xi_{k+1}}[\Lambda_{k+1}(s)]]\\
    &={\rm Tr}_{F_{k}}[\Lambda_k(s)^{-\frac{1}{2}}{\rm Tr}_{\Xi_{k+1}}[X_{F_{k+1}}]\Lambda_k(s)^{-\frac{1}{2}}\Lambda_k(s)]\\
    &={\rm Tr}_{F_{k+1}}[X_{F_{k+1}}]~.
\end{split}
\end{equation}
and the adjoint of $\mathcal{M}_k$ is obtained by inverting the Hilbert-Schmidt product as:
    \begin{equation}
        \begin{split}
            (\mathcal{M}_k^\dagger(x),X'_{F_{k+1}})_{F_{k+1}}&=(x,\mathcal{M}_k(X'_{F_{k+1}}))_\Gamma \\
            &={\rm Tr}_\Gamma[xs^{\frac{1}{2}}\Lambda_{k+1}^\dagger(\mathbb{I}_{\Xi_{k+1}}\otimes\Lambda_k(s)^{-\frac{1}{2}}{\rm Tr}_{\Xi_{k+1}}[X'_{F_{k+1}}]\Lambda_k(s)^{-\frac{1}{2}})s^{\frac{1}{2}}]\\
            &={\rm Tr}_{F_{k+1}}[\Lambda_{k+1}(s^{\frac{1}{2}}xs^{\frac{1}{2}})\mathbb{I}_{\Xi_{k+1}}\otimes\Lambda_k(s)^{-\frac{1}{2}}{\rm Tr}_{\Xi_{k+1}}[X'_{F_{k+1}}]\Lambda_k(s)^{-\frac{1}{2}}]\\
            &={\rm Tr}_{F_k}[{\rm Tr}_{\Xi_{k+1}}[\Lambda_{k+1}(s^{\frac{1}{2}}xs^{\frac{1}{2}})]\Lambda_k(s)^{-\frac{1}{2}}{\rm Tr}_{\Xi_{k+1}}[X'_{F_{k+1}}]\Lambda_k(s)^{-\frac{1}{2}}]\\
            &={\rm Tr}_{F_k}[\Lambda_{k}[s^{\frac{1}{2}}xs^{\frac{1}{2}}]\Lambda_k(s)^{-\frac{1}{2}}{\rm Tr}_{\Xi_{k+1}}(X'_{F_{k+1}})\Lambda_k(s)^{-\frac{1}{2}}]\\
            &={\rm Tr}_{F_{k+1}}[\mathbb{I}_{\Xi_{k+1}}\otimes\Lambda_k(s)^{-\frac{1}{2}}\Lambda_{k}[s^{\frac{1}{2}}xs^{\frac{1}{2}}]\Lambda_k(s)^{-\frac{1}{2}}X'_{F_{k+1}}]~,
        \end{split}
    \end{equation} 
thus giving
\begin{equation}
    \mathcal{M}_k^\dagger(x)=\mathbb{I}_{\Xi_{k+1}}\otimes\Lambda_k(s)^{-\frac{1}{2}}\Lambda_{k}(s^{\frac{1}{2}}xs^{\frac{1}{2}})\Lambda_k(s)^{-\frac{1}{2}}~.
\end{equation}

\subsection{$\mathcal{G}_k$ and its properties}
We begin by recalling the form of $\mathcal{M}_k$:
\begin{equation}
    \mathcal{M}_k(X_{k+1})\equiv s^{\frac{1}{2}}\Lambda_{k+1}^\dagger(\mathbb{I}_{\Xi_{k+1}}\otimes\Lambda_k(s)^{-\frac{1}{2}}{\rm Tr}_{\Xi_{k+1}}[X_{k+1}]\Lambda_k(s)^{-\frac{1}{2}})s^{\frac{1}{2}}
\end{equation}
and of the $(k+1)$-th Petz map:
\begin{equation}
    \mathcal{P}_{\Lambda_{k+1}}^{s}(X_{F_{k+1}})= s^{\frac{1}{2}}\Lambda_{k+1}^{\dagger}[\Lambda_{k+1}(s)^{-\frac{1}{2}}X_{F_{k+1}}\Lambda_{k+1}(s)^{-\frac{1}{2}}]s^{\frac{1}{2}}~.
\end{equation}
We write $\mathcal{G}_k:\mathcal{D}(\Gamma)\to\mathcal{D}(\Gamma)$ in terms of a supplementary map $\mathcal{S}_k:\mathcal{D}(\Gamma)\to\mathcal{D}(F_{k+1})$ as
\begin{equation}
    \mathcal{G}_k(x)=s^{\frac{1}{2}}\Lambda_{k+1}^\dagger(\mathcal{S}_k(x))s^{\frac{1}{2}}~,
\end{equation}
with $x\in\mathcal{D}(\Gamma)$, where $\mathcal{S}_k$ acts on the image of $\mathcal{P}_{\Lambda_{k+1}}^{s}$ as
\begin{equation}
    \mathcal{S}_k[\mathcal{P}_{\Lambda_{k+1}}^{s}(X_{F_{k+1}})]=\mathbb{I}_{\Xi_{k+1}}\otimes\Lambda_k(s)^{-\frac{1}{2}}{\rm Tr}_{\Xi_{k+1}}[X_{F_{k+1}}]\Lambda_k(s)^{-\frac{1}{2}}~.
    \label{a.Supp_map}
\end{equation}

Next, we show $\mathcal{G}_k$ is TP on ${\rm Im}(\mathcal{P}^s_{\Lambda_{k+1}}\circ\Lambda_{k+1})$. First, we notice the composition $\mathcal{P}^s_{\Lambda_{k}}\circ\Lambda_k$ is CPTP for all $k$ by construction. Using that $\mathcal{P}^s_{\Lambda_{k}}\circ\Lambda_k=\mathcal{G}_k\circ\mathcal{P}^s_{\Lambda_{k+1}}\circ\Lambda_{k+1}$, we get 
\begin{equation}
    {\rm Tr}[\mathcal{P}^s_{\Lambda_{k+1}}\circ\Lambda_{k+1}(x)]={\rm Tr}[x]={\rm Tr}[\mathcal{P}^s_{\Lambda_{k}}\circ\Lambda_k(x)]={\rm Tr}[\mathcal{G}_k\circ\mathcal{P}^s_{\Lambda_{k+1}}\circ\Lambda_{k+1}(x)]
\end{equation}
and thus that $\mathcal{G}_k$ is also TP on ${\rm Im}(\mathcal{P}^s_{\Lambda_{k+1}}\circ\Lambda_{k+1})\subseteq{\rm Im}(\mathcal{P}^s_{\Lambda_{k}})$.

\subsection{Proof of Eq.~\eqref{e.Petz_D_proof}% and \eqref{e.Gamma from F_k}
}
\label{a.Gamma from F_k}
Let us begin by re-expressing the quantum Markov chain \eqref{e.RP_via_cond_mutual_info} via the identity
\begin{equation}
    \mathcal{I}(A:B)=S(\rho_{A\cup B}||\rho_A\otimes\rho_B)
\end{equation}
as
\begin{equation}
S(\rho_{\Gamma\cup F_{k+1}}||\rho_\Gamma\otimes\rho_{F_{k+1}})=S(\rho_{\Gamma\cup F_k}||\rho_\Gamma\otimes\rho_{F_k})
\label{e.equal_cond_entro}
\end{equation}
where $\rho_P={\rm Tr}_{(\Gamma\cup\Xi)\setminus P}[\hat{U}(x_\Gamma\otimes  X_\Xi)\hat{U}^\dagger]$ for all $P$ appearing, and where we used the usual notation with all density operators denoted by lowercase Greek letters for the sake of simplicity. The above has a direct interpretation in terms of the Petz recovery. As we reviewed in Sec.~\ref{s.recovery}, Eq.~\eqref{e.equal_cond_entro} is equivalent to the existence of a perfect CPTP recovery map via the Petz theorem: if $S(\rho||\sigma)=S(\Lambda(\rho)||\Lambda(\sigma))$, then said recovery map exists and has the form \eqref{e.Petz}. Taking $\rho=\rho_{\Gamma\cup F_{k+1}}$, $\sigma=\rho_{\Gamma}\otimes\rho_{F_{k+1}}$, and $\Lambda={\rm Tr}_{\Xi_{k+1}}$ (and thus $\Lambda^\dagger(\rho)=\rho\otimes\mathbb{I}_{\Xi_{k+1}}$) in the definition of the Petz map \eqref{e.Petz} gives
\begin{equation}
    \mathcal{P}^\sigma_{{\rm Tr}_{\Xi_{k+1}}}(\rho)=(\rho_\Gamma\otimes\rho_{F_ {k+1}})^{1/2}\mathbb{I}_{\Xi_{k+1}}\otimes((\rho_\Gamma\otimes\rho_{F_ {k}})^{-1/2}\rho (\rho_\Gamma\otimes\rho_{F_ {k}})^{-1/2})(\rho_\Gamma\otimes\rho_{F_ {k+1}})^{1/2}~,
\end{equation}
and apply it to a state written in the tensor product decomposition $\rho=\sum_ir_i x_i\otimes X_{F_k}^i$ to get:
    \begin{equation}
    \mathcal{P}^\sigma_{{\rm Tr}_{\Xi_{k+1}}}(\rho_{\Gamma\cup F_k})=\sum_i r_i x_i\otimes\left[\rho_{F_{k+1}}^{\frac{1}{2}}(\rho_{F_{k}}^{-\frac{1}{2}}X_{F_k}^i\rho_{F_{k}}^{-\frac{1}{2}}\otimes\mathbb{I}_{\Xi_{k+1}})\rho_{F_{k+1}}^{\frac{1}{2}}\right]~,
    \label{e.Redundancy_petz}
\end{equation}
and thus \eqref{e.Petz_D_proof} holds, with
\begin{equation}
    \mathcal{R}_{F_k\rightarrow F_{k+1}}(X)=\rho_{F_{k+1}}^{\frac{1}{2}}(\rho_{F_{k}}^{-\frac{1}{2}}X\rho_{F_{k}}^{-\frac{1}{2}}\otimes\mathbb{I}_{\Xi_{k+1}})\rho_{F_{k+1}}^{\frac{1}{2}}~.
\end{equation}

\end{document}